%% file: p.tex
\pdfoutput=1

\DocumentMetadata{}
\documentclass[sigplan]{acmart}
\acmSubmissionID{1264}
\renewcommand\footnotetextcopyrightpermission[1]{}
\include{pkgs}
\usepackage{bm}

\newcommand{\sys}{\mbox{\textsc{TeeMate}}\xspace}

\newcommand{\VAzero}{\mbox{$\cc{VA}_0$}\xspace}
\newcommand{\PAzero}{\mbox{$\cc{PA}_0$}\xspace}

\newcommand{\VAone}{\mbox{$\cc{VA}_1$}\xspace}
\newcommand{\PAone}{\mbox{$\cc{PA}_1$}\xspace}

\newcommand{\native}{\mbox{\textbf{\cc{Native}}}\xspace}
\newcommand{\strawman}{\mbox{\textbf{\cc{Strawman}}}\xspace}

\newcommand{\enclaveAliasing}[0]{\mbox{EPC aliasing}\xspace}

\input{cmds}
\begin{document}

\sloppy
\input{hdr}

\input{abstract}

\maketitle
\pagestyle{plain}

\input{intro}

\input{back}

\input{motiv}

\input{key}

\input{design}

\input{impl}

\input{eval}

\input{secanalysis}

\input{discuss}

\input{relwk}

\input{conclusion}

\bibliographystyle{plain}
\bibliography{p,sslab,conf}
\end{document}

%% file: pkgs.tex
\usepackage{chngpage}

\usepackage{amsmath,amsopn,amssymb}
\usepackage{subcaption}
\usepackage{endnotes,microtype,xspace,graphicx,fancyvrb,multirow}
\usepackage{booktabs}
\usepackage{array,underscore,relsize}
\usepackage{fancyhdr}
\usepackage{enumitem}
\usepackage[labelfont=bf,font=small,skip=5pt]{caption}
\usepackage{fp}
\usepackage{siunitx}

\usepackage{balance}

\usepackage{pifont}

%% file: cmds.tex
\newcommand{\cc}[1]{\mbox{\smaller[0.5]\texttt{#1}}}

\fvset{fontsize=\scriptsize,xleftmargin=8pt,numbers=left,numbersep=5pt}

\input{code/fmt}

\def\Snospace~{\S{}}

\newcommand{\x}{$\times$\xspace}

\newif\ifdraft\drafttrue
\newif\ifnotes\notestrue
\ifdraft\else\notesfalse\fi

\input{glyphtounicode}
\newcolumntype{R}[1]{>{\raggedleft\let\newline\\\arraybackslash\hspace{0pt}}p{#1}}

\newcommand{\squishlist}{
\begin{itemize}[noitemsep,nolistsep]
  \setlength{\itemsep}{-0pt}
}
\newcommand{\squishend}{
  \end{itemize}
}

\usepackage{tikz}
\usetikzlibrary{fit}
\newcommand*\WC[1]{%
\begin{tikzpicture}[baseline=(C.base)]
\node[draw,circle,inner sep=0.2pt](C) {#1};
\end{tikzpicture}}

\usepackage{xstring}
\newcommand{\PP}[1]{
\vspace{2px}
\noindent{\bf \IfEndWith{#1}{.}{#1}{#1.}}
}

\newcommand{\boxbeg}{
\vspace{2px}
\noindent\begin{tabular}{|l|}\hline
\begin{minipage}{3.2in}
\vspace{2px}
\noindent
}

\newcommand{\boxend}{
\vspace{2px}
\end{minipage}\\ \hline
\end{tabular}
\vspace{-10pt}
}

%% file: hdr.tex
\title{\sys: Fast and Efficient Confidential Container using Shared Enclave}

\author{Chulmin Lee\textsuperscript{$\dagger$}}
\affiliation{%
\institution{Seoul National University}
\country{}
}
\email{2harry@snu.ac.kr}

\author{Jaewon Hur\textsuperscript{$\dagger$}}
\affiliation{%
\institution{Seoul National University}
\country{}
}
\email{hurjaewon@snu.ac.kr}

\thanks{$\dagger$ These authors contributed equally to this work.}

\author{Sangho Lee}
\affiliation{%
\institution{Microsoft Research}
\country{}
}
\email{Sangho.Lee@microsoft.com}

\author{Byoungyoung Lee}
\affiliation{%
\institution{Seoul National University}
\country{}
}
\email{byoungyoung@snu.ac.kr}

%% file: abstract.tex
\begin{abstract}
  Confidential container is becoming increasingly popular as
  it meets both needs for efficient resource management by
  cloud providers, and data protection by cloud users.
  Specifically, confidential containers integrate the
  container and the enclave, aiming to inherit the
  design-wise advantages of both---i.e., resource management
  and data protection.
  However, current confidential containers suffer from large
  performance overheads caused by i)~a larger startup
  latency due to the enclave creation, and ii)~a larger
  memory footprint due to the non-shareable characteristics
  of enclave memory.

  This paper explores a design conundrum of confidential
  container, examining why the confidential containers
  impose such large performance overheads.
  Surprisingly, we found there is a universal misconception
  that an enclave can only be used by a single
  (containerized) process that created it.
  However, an enclave can be shared across multiple
  processes, because an enclave is merely a set of physical
  resources while the process is an abstraction constructed
  by the host kernel.

  To this end, we introduce \sys, a new approach to utilize
  the enclaves on the host system.
  Especially, \sys designs the primitives to i)~share the
  enclave memory between processes, thus preserving memory
  abstraction, and ii)~assign the threads in enclave between
  processes, thus preserving thread abstraction.
  We concretized \sys on Intel SGX, and implemented
  confidential serverless computing and confidential
  database on top of \sys based confidential containers.
  The evaluation clearly demonstrated the strong practical
  impact of \sys by achieving at least 4.5 times lower
  latency and 2.8 times lower memory usage compared to the
  applications built on the conventional confidential
  containers.

\end{abstract}

%% file: intro.tex
\section{Introduction}
\label{s:intro}

Cloud computing offers several advantages in resource management, allowing its
users to focus on their application development without the burdens of managing
computing resources~\cite{awslambda, azurefunctions, amazonsaas, azuresaas,
kaas,kubernetes}.
Specifically, cloud providers take the complete charge of
managing the entire system resources (e.g., CPU, memory, and
storage) on which cloud users easily run their application.
Due to these advantages, emerging cloud service models such
as Software as a Service (SaaS)~\cite{amazonsaas, azuresaas}
and Kubernetes as a Service (KaaS)~\cite{kaas} have gained
the strong popularity.

Looking into the technical aspect of these cloud service
models, \emph{container}~\cite{docker,lxc} (or OS-level
virtualization) plays the key role as it facilitates both
resource management and isolation.
To be specific, each container is assigned with isolated resources,
allowing the cloud providers to manage the resources per process,
avoiding costly full virtualization~\cite{containervsVM}.
This is enabled by two key features provided by the underlying OS:
i)~namespace~\cite{namespace}, which provides a different userland view over
the resources per process (e.g., files and network), and
ii)~cgroup~\cite{cgroup}, which limits the CPU and memory usage per process.
These two features construct a containerized environment,
which serves as a basic management unit by the cloud
providers.

Meanwhile, confidential
computing~\cite{gcore-cc,azure-cc,cloudsigma-cc,sev,cca} has gained
strong popularity in clouds, as cloud users demand strong security
guarantees over their data.
Especially, there is a growing need to exclude the cloud
providers from trusted path, as the cloud handles a large
amount of privacy-sensitive data that can be in
conflict of interest.
To meet such security demands, confidential computing
(including Intel SGX~\cite{gcore-cc}, AMD SEV~\cite{sev},
and Intel TDX~\cite{inteltdx}) introduces an \emph{enclave},
which is a trusted execution environment.
Specifically, the enclave is protected from all systems components
including operating systems, hypervisors, and even the other
enclaves~\cite{costan2016intel,sev} such that the enclave
owner can safely execute their workloads without trusting
the cloud providers.

In this paper, we explore a design conundrum of \emph{confidential
  containers}~\cite{arnautov2016scone}, which integrates
aforementioned two techniques, containers and confidential computing.
To be specific, confidential containers aim at inheriting the
design-wise advantages of each---i.e., resource management capability
from containers and data protection capability from confidential
computing.
As it is naturally thought, current confidential containers construct
a single container with a single enclave, which serves as a basic
management unit by the host system.
For example, SGX enclaves are already widely used with the
containers to construct a protected environment in the
container~\cite{gcore-cc}.

However, current confidential containers suffer from large
performance overheads.
Especially, we found that integrating the enclave into the
container incurs
i)~a larger startup latency due to the inherent security
mechanism when creating the
enclaves~\cite{li2021confidential,shixuan2023reusableenclave},
and
ii)~a larger memory footprint due to the non-shareable
characteristics of enclave memory~\cite{li2021confidential,jason2022elasticlave,lee2022cerberus}.
We confirmed these overheads and root causes by conducting
preliminary experiments on two popular cloud applications,
i.e., serverless computing~\cite{awslambda,azurefunctions,openwhisk}, and database~\cite{redis,postgresql}.

Surprisingly, after the analysis, we found there is a
universal misconception that a single enclave must be
dedicated to only a single process.
In other words, all the previous works, which use the
enclaves~\cite{shen2020occlum,ahmad2021chancel,arnautov2016scone,tsai2017graphene,Baumann2014haven,hunt2018ryoan,shinde2017panoply}
have (incorrectly) assumed the process that initially created the enclave can
exclusively own it.
However, we found that it is wrong, and an enclave can be
shared across multiple processes (and containers) as long as
we preserve the memory abstraction and thread abstraction
assumed by the host operating systems.

Based on this observation, we design \sys, which is a new
approach to utilize the enclaves in the host's perspective.
In particular, \sys enables a single enclave to be shared
across multiple processes so that the host kernel can manage
(and isolate) them with different containers.
Thus, \sys successfully solves the issues of startup latency
and memory footprint as it avoids creating a new enclave
every time, and enables to share the enclave memory between
different (containerized) processes on the host kernel.

In order to share the same enclave across different
processes, we design \sys to
i)~preserve the memory abstraction on the memory of shared
enclave, and
ii)~preserve the thread abstraction on the enclave's
threads.
Then, we discuss intra-enclave isolation to guarantee the
isolation between the processes using the same enclave.
We concretized these concepts in the context of Intel SGX,
which is an already widely used confidential computing
technology in the cloud~\cite{gcore-cc}.

We implemented \sys on the secured version of two major
cloud applications,
i)~confidential serverless computing, and
ii)~confidential database.
The evaluation results clearly indicate that \sys outperforms applications using
current confidential containers in terms of latency and memory usage.
Specifically, \sys achieves a latency speedup of 4.54-6.98 times and exhibits
memory footprint by only 20-36\% in confidential serverless computing.
Similarly, in confidential database applications, \sys achieves a latency
speedup of 277.6-1046.6 times and reduces memory footprint up to 41\%.
Thus, we confirm that \sys is a ready to use framework on
current cloud infrastructure, which solves the performance
issues without any hardware modification.

%% file: back.tex
\section{Background}
\label{s:back}

This section explains container
(\autoref{ss:back-container}) and enclave
(\autoref{ss:back-enclave}), which are the basic unit of
resource management and data protection.
Then, we introduce confidential container, which integrates
both to inherit the design-wise advantages
(\autoref{ss:back-coco}).

\subsection{Container based Resource Management}
\label{ss:back-container}

Container technologies~\cite{docker,lxc} refer to the
software for managing and isolating the system resources
using OS primitives~\cite{namespace,cgroup}.
Especially, the containers work at the granularity of a process, as it is the
basic unit for organizing the system resources (e.g., virtual address space, CPU
registers, and opened files per each process).
In particular, containers are implemented using kernel
subsystems,
i)~namespace~\cite{namespace}, which provides a different
(isolated) userland view over the system resources (e.g.,
files and network), and
ii)~cgroup~\cite{cgroup}, which limits the CPU and memory used by
the process.
They populate a containerized environment, inside which the
process runs as if it has its own computer system.

Cloud providers heavily rely on the containers as they
facilitate the resource management and isolation with
minimal performance overheads~\cite{googlecloudcontainer,amazon-ecs}.
Specifically, recent trends in cloud industry have triggered
the widespread adoption of containers such as serverless
computing~\cite{awslambda,azurefunctions} and micro-service
architecture~\cite{microservice}.
For example, almost 60\% of organizations that use the cloud
have adopted the container
technologies~\cite{container-adoption}.
As another example, Software as a Service (SaaS) model,
which also heavily uses the containers, is expected to grow
at a CAGR of 13\% from 2023 to 2030~\cite{saas-cagr}.

\subsection{Enclave based Data Protection}
\label{ss:back-enclave}

Enclave is the basic unit of protection by confidential
computing technologies (e.g., Intel SGX~\cite{gcore-cc}, AMD
SEV~\cite{sev}, and Intel TDX~\cite{inteltdx}) that is
hardware isolated from the systems components (e.g.,
operating systems and hypervisors).
To be specific, the CPUs construct an enclave by
i)~populating and encrypting the memory region used by the
enclave, and
ii)~protecting the enclave's register context through access
control.
More specifically, when creating an enclave, an initial
image (i.e., code and data) is copied into the protected
memory region, and hash checked so that the enclave owner
can ensure the integrity of the loaded image.
After that, CPU encrypts the memory and isolates the
register context such that any other components including
even another enclave cannot access the original one's data,
thus guaranteeing the confidentiality.

Thanks to its strong security guarantees, emerging cloud
applications use the enclave to protect their data in a
potentially compromised cloud environment.
Especially, current trends of AI to handle a large amount of
privacy-sensitive data have pushed to use the enclaves.
In response, several open source projects for confidential
computing have been initiated
(e.g.,~\cite{CNCFCOCO,amd-sevese}) and cloud providers
quickly announced the support for confidential computing
(e.g.,~\cite{amazonsovereign,azure-cc}).

\subsection{Confidential Container: Intersection of
  Container and Enclave}
\label{ss:back-coco}

Confidential container~\cite{arnautov2016scone,CNCFCOCO} has
been introduced to meet both needs in the cloud industry as
explained above.
To be specific, confidential container integrates the
container and the enclave to inherit the design-wise
advantages of both---i.e., efficient resource management by
cloud providers, and data protection for cloud users.
For example, SGX enclaves are already widely used for the
confidential containers as they are originally designed for
process level isolation~\cite{costan2016intel}.
In addition, confidential virtual machines (VMs), which use
AMD SEV~\cite{sev}, or Intel TDX~\cite{inteltdx}, are also actively
studied to be integrated with the containers (e.g., Kata
Container~\cite{katacontainers}).
Regardless of the underlying technology, current
confidential containers assign an enclave for each
container, which naturally follows the concept of the
container and the enclave.

%% file: motiv.tex
\section{Infeasibility of Confidential Container}
\label{s:motiv-inf}

However, current confidential containers suffer from large
performance overheads~\cite{li2021confidential,shixuan2023reusableenclave,li2021confidential,jason2022elasticlave,lee2022cerberus}.
Especially, integrating the enclave with the container
imposes
i)~a longer startup latency due to the creation of the
enclave, and
ii)~a larger memory footprint due to the non-shareable
characteristics of the enclave's memory.
In order to clearly demonstrate these points, we conducted
preliminary experiments as follows.

\subsection{Preliminary Experiments on the Performance
  Overheads}
\label{ss:motiv-inf-pre}
We designed the preliminary experiments with two following
research questions:

\vspace{2pt}
\begin{enumerate}[nolistsep,leftmargin=1.5em]
  \setlength\itemsep{1pt}
  \setlength\parskip{1pt}
  \item How much performance overheads does confidential
        container impose?
  \item Why does confidential container impose such
        performance overheads?
\end{enumerate}
\vspace{5pt}

To this end, we measured the startup latency and memory
footprint of two benchmark applications, which are expected
to be widely used with confidential containers:
i)~confidential serverless computing, and
ii)~confidential database.

\begin{figure}[t]
  \centering
  \begin{subfigure}{0.48\columnwidth}
    \includegraphics[width=\linewidth]{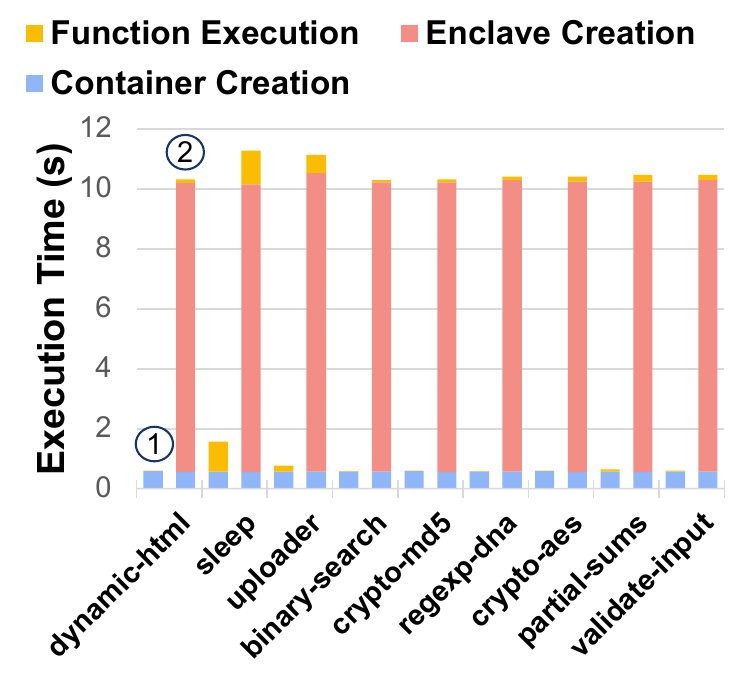}
    \caption{Latency for handling a request.}
  \label{fig:motiv-serverless-latency}
  \end{subfigure}
  \begin{subfigure}{0.48\columnwidth}
    \includegraphics[width=\linewidth]{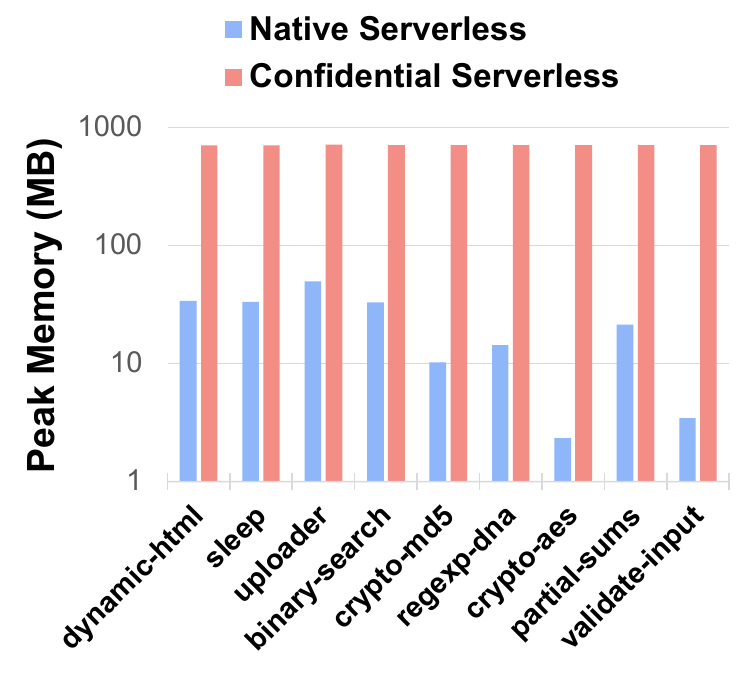}
    \caption{Memory usage for a request.}
  \end{subfigure}
  \label{fig:motiv-serverless-memory}
  \caption{Performance comparison of confidential serverless
    computing versus the native model.
    \protect\WC{1}: \cc{Native Serverless}, \protect\WC{2}:
    \cc{Confidential Serverless}.
    The bloated memory of confidential serverless computing is due to the
    Gramine LibOS's implementation that physically populates all the allocated
    virtual memory~\cite{tsai2017graphene}.}
  \label{fig:motiv-serverless}
\end{figure}
\PP{Performance Overheads of Confidential Serverless
  Computing}
Serverless computing is an emerging cloud computing model,
where the resource management is fully delegated to the
cloud providers while the users can solely focus on their
workloads~\cite{awslambda,azurefunctions}.
Containers~\cite{docker,lxc} are the key building blocks in such model
as they facilitate the resource management and isolation by
the cloud providers.
In particular, the providers construct a different
containerized environment for each request and the following
function instance (i.e., unit of the computation in
serverless computing~\cite{openwhisk}), thereby providing a
different userland view of resources and resource limits.

Confidential serverless
computing~\cite{shixuan2023reusableenclave,li2021confidential},
which employs the confidential containers, is the security
enhanced version of serverless computing as it protects the
workloads even on a compromised cloud environment.
Meanwhile, the providers still manage the system resources
so that the users can only focus on their workloads.
To this end, state-of-the-art confidential serverless
computing frameworks serve each request by creating a new
confidential container (including a new containerized
process and a new enclave), and running a function instance
on it.

Thus, we measured the latency for handling each request and
memory footprint of the state-of-the-art confidential
serverless computing framework.
To be specific, we implemented a security enhanced version
of OpenWhisk~\cite{openwhisk} that runs the functions in the
confidential containers, using an SGX
enclave~\cite{gcore-cc} and Gramine LibOS~\cite{gramine} as
an enclave runtime.
As shown in~\autoref{fig:motiv-serverless}, employing current confidential
container in serverless computing imposes almost 10\x latency slowdown and
20-300\x more memory usage.
Especially, \autoref{fig:motiv-serverless-latency} shows that
creating an enclave (for every request) takes 16.8-17.4\x longer
time than creating a containerized process, demonstrating
that creating an enclave (and its security mechanisms) is a
major bottleneck.

\begin{figure}[t]
  \centering
  \begin{subfigure}{0.48\columnwidth}
    \includegraphics[width=\linewidth]{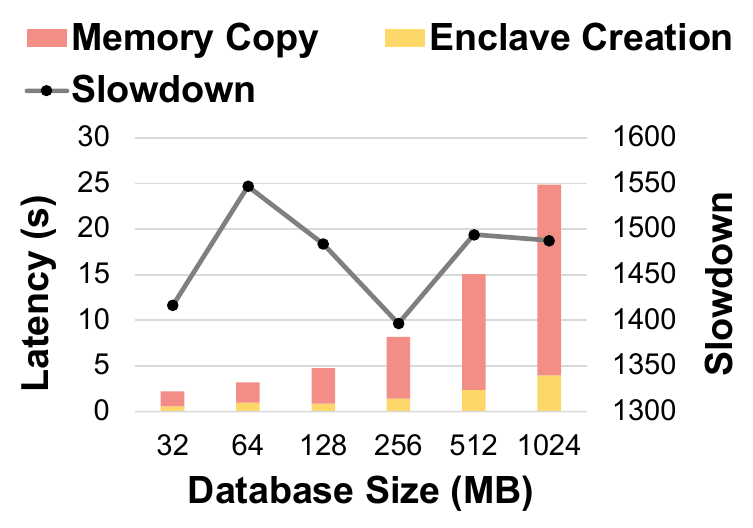}
    \caption{Latency of forking a child process for snapshot.}
  \end{subfigure}
  \hspace{5pt}
  \begin{subfigure}{0.48\columnwidth}
    \vspace{8pt}
    \includegraphics[width=\linewidth]{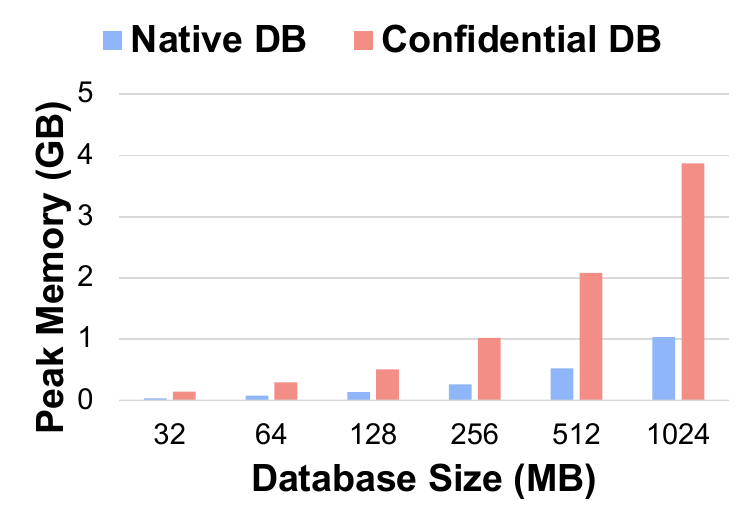}
    \caption{Memory usage of database system with fork-based
      snapshot.}
  \end{subfigure}
  \caption{Performance comparison of confidential database
    system versus the native system.}
  \label{fig:motiv-database}
\end{figure}

\PP{Performance Overheads of Confidential Database}
Database is one of the most widely used applications in
cloud computing~\cite{postgresql,redis}.
In particular, database as a service (DBaaS)~\cite{dbaas} is a
standard way to quickly integrate the database into the
user's service logic while the cloud providers manage the
underlying software stacks.
DBaaS also heavily relies on the containers as the providers
can easily manage and isolate the resources between
different service instances.

Especially, the file systems management of the containers
facilitates the fork-based snapshots for database
systems~\cite{redisFork}.
In this approach, the parent database process continues to
handle the requests by forking a child process, while the
child process performs the snapshot by writing the database
into the storage.
Fork-based snapshot is a widely used approach by database
systems~\cite{redisFork}, as it provides a significant performance
benefits in accordance with the copy-on-write semantics of
operating systems~\cite{Copy-on-write}---i.e., data pages are 
copied only when a new request to write to that page is received.

Confidential database, which runs the database system in the
confidential containers, protects sensitive data from the
cloud providers.
With current confidential containers, fork-based snapshot
for database would create a new (child) process and enclave,
copy the parent enclave's memory into the child's enclave
(following the fork's semantics~\cite{fork}), and run as
usual---i.e., the parent serving the requests while the
child performing the snapshot.
However, copy-on-write semantics would not be allowed as the
parent's enclave and the child's enclave cannot share the
same protected memory~\cite{shen2020occlum}.

Thus, we measured the startup latency of fork-based snapshot
and memory footprint of current confidential database
systems.
To be specific, we implemented a security enhanced version
of Redis~\cite{redis} that runs in the confidential
container, which employs an SGX enclave~\cite{gcore-cc} and
Gramine LibOS~\cite{gramine}.
As shown in~\autoref{fig:motiv-database}, forking a child
process (and copying the enclave) takes over 1000$\times$ times
longer latency and 4$\times$ times larger memory usage as the
entire memory contents of the parent enclave need to be
copied to the child enclave.
While the numbers may be exaggerated by the implementation
of Gramine LibOS (i.e., copying memory through TLS encrypted
channel~\cite{tls}), we want to note that inability to share
the memory between parent and child enclave is the main
cause of these overheads.

\begin{table}[t]
  \centering
  \captionof{table}{Previous works to solve the performance
    issues of enclave.}
  \input{fig/motiv/table-prev}
  \label{tab:motiv-prev}
\end{table}

\subsection{Limitations of Previous Works to Solve the
  Performance Overheads}
\label{ss:motiv-inf-prev}
Several previous works have tried to solve the performance
issues of enclaves as illustrated
in~\autoref{tab:motiv-prev}.
In particular, we categorize them into two lines of works as
follows:
i)~improving the startup latency of enclave, and
ii)~enabling the efficient memory sharing between enclaves.

However, most of the works cannot be directly applied to
current cloud platforms~\cite{amazonsaas,azuresaas,kaas,googlecloudcontainer,oracle-cloud,ibmcloudfunctions}, 
as they require hardware modification (i.e., shown 
in~\autoref{tab:motiv-prev}).
While Reusable Enclave~\cite{shixuan2023reusableenclave} 
achieves the goal without modifying the hardware, it enforces
the cloud providers to maintain the same container to reuse 
the enclave, thereby making it difficult to manage the resources.
Specifically, all of them have only focused on the inherent
issues of enclaves, but not on how to utilize the enclaves
in the perspective of host systems.

%% file: fig/motiv/table-prev.tex
\resizebox{0.9\linewidth}{!}{
  \begin{tabular}{c|cc}
    \toprule
    Research goal & Scheme & Hardware modified \\
    \midrule
    \multirow[b]{2}{*}{Fast enclave} & PENGLAI~\cite{feng2021scalable} & $\bigcirc$ \\
                  & PIE~\cite{li2021confidential} & $\bigcirc$ \\
    \multirow[t]{2}{*}{startup} & LightEnclave~\cite{gu2022lightenclave} & $\bigcirc$ \\
                  & Reusable Enclave~\cite{shixuan2023reusableenclave} & $\times$\textsuperscript{$\dagger$} \\
    \midrule
    \multirow[b]{2}{*}{Efficient enclave} & Nested Enclave~\cite{park2020nested} & $\bigcirc$ \\
                  & PIE~\cite{li2021confidential} & $\bigcirc$ \\
    \multirow[t]{2}{*}{memory sharing} & Elasticlave~\cite{jason2022elasticlave} & $\bigcirc$ \\
                  & Cerberus~\cite{lee2022cerberus} & $\bigcirc$ \\
    \bottomrule
    \multicolumn{3}{l}{\small{\textsuperscript{$\dagger$}Reusable Enclave discusses only for temporally reusing an enclave.}}
  \end{tabular}
}

%% file: key.tex
\section{Key Idea of \sys}
\label{s:key}

Thus, we re-thought the enclave in the perspective of host
systems, and surprisingly, we found an incorrect universal
assumption on its usage (\autoref{ss:key-mis}).
Based on this observation, we came up with the key idea of
\sys, which solves the aforementioned issues
(\autoref{ss:key-our}).

\begin{figure}
  \centering
  \begin{subfigure}{0.45\columnwidth}
    \includegraphics[width=\linewidth]{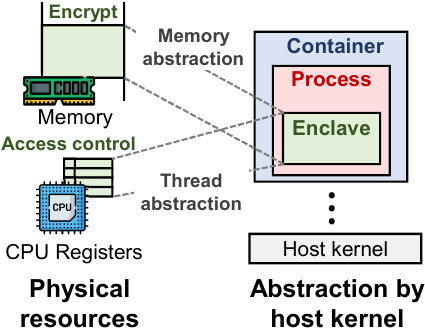}
    \caption{Previous assumption on enclave.
      One-to-one enforcement of a process and an enclave.}
  \end{subfigure}
  \hspace{10pt}
  \begin{subfigure}{0.45\columnwidth}
    \includegraphics[width=\linewidth]{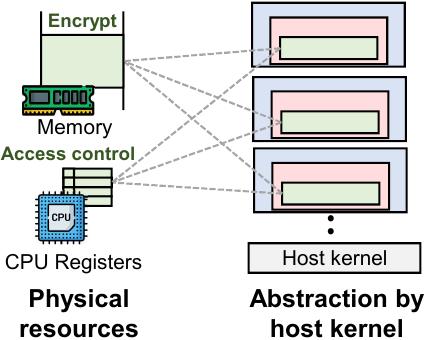}
    \caption{Key idea of \sys.
      Sharing an enclave across multiple processes.}
  \end{subfigure}
  \caption{Relation between an enclave and process in the
    perspective of host kernel.}
  \label{fig:motiv-relation}
\end{figure}

\subsection{Universal Assumption: One-to-One Enforcement
  of Process and Enclave}
\label{ss:key-mis}

After analyzing the issues, we found there is a universal
assumption that a single enclave must be dedicated to only a
single process (i.e., one-to-one enforcement of a process
and an enclave as shown
in~\autoref{fig:motiv-relation}-(a)).
For example of using SGX enclaves, no previous work has
assumed using the same enclave by different
processes~\cite{feng2021scalable,li2021confidential,gu2022lightenclave,shixuan2023reusableenclave}, 
and they implicitly assumed the process which creates the enclave 
exclusively owns it.
While Occlum~\cite{shen2020occlum} designs multiprocessing
in a single enclave, it is not about how the processes (of
the host kernel) use the enclave---i.e., Occlum is also a
single process in the perspective of the host kernel.

However, we found that this assumption is wrong.
In other words, a single enclave does not have to be
dedicated to a single process.
This is because an enclave is merely a protected resource
composed of an encrypted memory and isolated CPU context,
while the process is an abstraction of the resources created
by the host kernel---i.e., the way of thinking the
resources.
In the perspective of host kernel, an enclave can be deemed
as any other resources, such as memory and disk, that can be
abstracted and shared across the processes.

\subsection{Our Solution: Sharing a Single Enclave across
  Multiple Containers}
\label{ss:key-our}

Based on this observation, we came up with the key idea of
\sys, sharing a single enclave across multiple containers
(i.e.,~\autoref{fig:motiv-relation}-(b)).
Since there is no need for an enclave to be dedicated to a
single process, it is also possible to share the same
enclave across multiple containerized processes.
By doing so, we can avoid the performance issues while
taking both benefits of the container and the enclave as the
conventional confidential containers.
In other words, cloud providers can efficiently manage the
system resources by applying different containerized
environment for each process, while the users can be ensured
the security of their data using the enclave.
However, we also preserve the performance, as we
i)~avoid creating a new enclave every time, and
ii)~enable sharing the enclave's memory between different
processes.

To this end, we design \sys to provide an abstraction for
the host kernel that it is operating different
(containerized) processes with dedicated enclaves, but
actually using the same enclave.
More specifically, we design the primitives to preserve
i)~the memory abstraction on the shared enclave's memory
(i.e.,~\autoref{ss:design-memory}), and
ii)~the thread abstraction on the shared enclave's threads
(i.e.,~\autoref{ss:design-thread}).
Then, we discuss how the isolation between the processes can
be achieved within the same enclave
(i.e.,~\autoref{ss:design-isolation}).

%% file: design.tex
\section{Design of \sys}
\label{s:design}

In this section, we introduce the threat model of \sys
(\autoref{ss:design-threat}), and explain the design of \sys
using Intel SGX~\cite{gcore-cc}
(\autoref{ss:design-overview} to
\autoref{ss:design-isolation}).

\begin{figure}[t]
  \centering
  \includegraphics[width=0.7\linewidth]{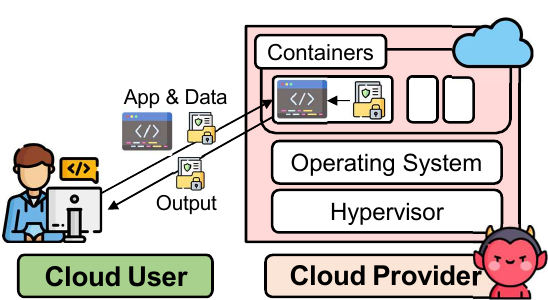}
  \caption{Threat model of \sys.}
\label{fig:design-threat}
\end{figure}

\subsection{Threat Model}
\label{ss:design-threat}
We assume the common threat model of confidential computing
as shown in~\autoref{fig:design-threat}, where the cloud
users do not trust the cloud providers.
This is because the cloud providers may be compromised, or
even themselves are in conflict of interests with the cloud
users (e.g., Samsung utilizing the services hosted by Amazon
AWS~\cite{samsung-amazon}).
We want to note that there is a growing demand to protect
the data even on a compromised cloud environment as more
privacy-senstive data are handled in the cloud.

As explained in~\autoref{ss:back-enclave}, confidential
computing is an emerging solution to meet such needs, so we
focus on the performance issues of employing both the
confidential computing and container technologies.
We do not consider general security issues of confidential
computing such as Iago attacks~\cite{checkoway2013iago} and
side
channels~\cite{kiriansky2018dawg,kurth2020netcat,lipp2018meltdown,kocher2020spectre}.
Denial-of-Service attacks~\cite{schuba1997analysis,dos} are also out-of-scope.
In addition, we trust the implementation of the software
components loaded in the enclave, and exploits through their
vulnerabilities are out-of-scope.
Hardening software implementations is a long been problem,
and we believe \sys can take advantages of ongoing
researches~\cite{cloosters2022sgxfuzz,wang2023symgx,gross2023fuzzilli}.

\subsection{Design Overview}
\label{ss:design-overview}
As illustrated in~\autoref{ss:key-our}, \sys enables high
performance confidential containers by sharing the same
enclave across different containers (i.e., containerized
processes).
To this end, \sys provide an abstraction for the host kernel
that it is operating different processes with dedicated
enclaves (within each containerized environment), but
actually using the same enclave (inside which, the resources
are isolated for each process).
Specifically, in order to achieve the process abstractions
assumed by the host kernel, \sys satisfies two requirements:
i)~for memory abstraction, sharing the same enclave's
physical memory within different virtual address spaces of
each process, and
ii)~for thread abstraction, assigning the threads in the
same enclave to each different process.

Thus, we design \sys's primitive operations achieving these
requirements based on Intel SGX~\cite{gcore-cc}.
While we provide the design for Intel SGX only, we want to
note the key idea of \sys (i.e., sharing the same enclave)
is general enough to be applied to confidential VMs (e.g.,
AMD SEV~\cite{sev}, Intel TDX~\cite{inteltdx}).
We discuss how \sys can be used for confidential VMs
in~\autoref{s:discuss}.

\begin{figure}[t]
  \centering
  \includegraphics[width=0.7\linewidth]{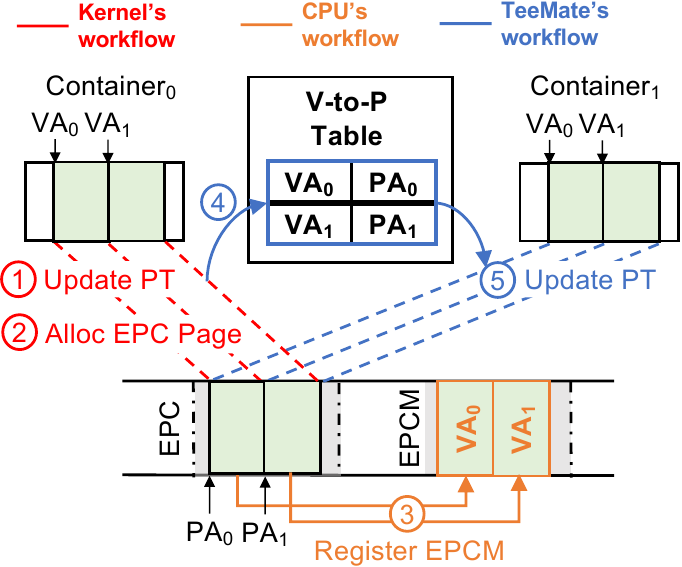}
  \caption{Sharing the same EPC pages across different containers
    using \enclaveAliasing}
\label{fig:design-VACondition}
\end{figure}

\subsection{Sharing Enclave Memory across Multiple
  Containers}
\label{ss:design-memory}

In order to provide the memory abstraction, \sys maps the
physical pages of the same enclave into the virtual address
spaces of each process.
Thus, \sys enables the threads running in different
containers (with different address spaces) to access the
same enclave's code and data, which avoids to create a new
enclave for every new container.
Especially in SGX, we name it Enclave Page Cache (EPC)
aliasing, as we alias the same EPC pages (of an SGX enclave)
to different virtual address spaces---i.e., EPC is a
protected memory region used by Intel SGX~\cite{gcore-cc}.
In the following, we explain the details how we alias the
EPC pages.

\PP{Technical Analysis: Address Translation and Validation
  in SGX}
SGX ensures the integrity of the virtual-to-physical address
mapping for EPC pages, preventing malicious systems
components from launching a page remapping attack (e.g.,
tricking a victim enclave to access a different EPC page
through the same virtual address~\cite{costan2016intel}).
This integrity is assured by maintaining additional address mapping
within the special EPC page, so called EPCM.
Specifically, when a new EPC page is allocated, the kernel updates the
page table with virtual-to-physical address mapping (e.g., \cc{VA} to
\cc{PA}) related to the EPC page.
When the kernel requests the CPU to create the EPC pages, the CPU
creates a new EPCM entry per new EPC page, which contains \cc{VA} of
the corresponding EPC page located at \cc{PA}.
When any access to the EPC page is attempted later using \cc{VA}, the
CPU translates \cc{VA} to \cc{PA} using the page table.
However, since the page table can be compromised by the
adversarial systems components, the CPU further validates
that such a translation is correct using EPCM, thereby
assuring the integrity of the virtual-to-physical address
mapping.

An interesting technical characteristic here is in the
address validation mechanism of EPCM, which does not involve
the identity of a process (i.e., EPCM does not include
process identifiers like ASID~\cite{armv7a}).
Thus, an EPC page (located at \cc{PA}) can be accessed by any
other processes using \cc{VA} as long as the page table of
the process contains the same mapping from \cc{VA} to
\cc{PA}.
If so, the process, which did not initially create the EPC page, can
still access the EPC page through accessing to the same \cc{VA}.
Note that while this may seem a vulnerable design, it does not harm
the security assurance of SGX, which we elaborate the detailed
security analysis in~\autoref{s:security}.

\PP{EPC Aliasing}
Based on this characteristic, \sys aliases EPC pages across
different processes.
Specifically, \enclaveAliasing maps multiple virtual pages to the same
physical EPC page, where multiple virtual pages are (i) associated
with different processes and (ii) these virtual pages have the same
virtual address.
More technically, if a process $\cc{p}_0$ already allocated an EPC
page with address mapping (i.e., \VAzero to \PAzero), \enclaveAliasing
allows to map the same EPC page to another process $\cc{p}_1$ by
updating $\cc{p}_1$'s page table, inserting \VAzero to \PAzero address
mapping.
After that, the process $\cc{p}_1$ can also access the EPC page
with virtual address \VAzero.

In order to share EPC pages between containers, \sys performs the
\enclaveAliasing as illustrated in~\autoref{fig:design-VACondition}.
Specifically, \sys creates the initial container.
Then, after identifying available virtual and physical pages for an
enclave, the kernel updates the page table of the initial container
---i.e., adding an address mapping from \VAzero and \VAone to
\PAzero and \PAone, respectively~(\WC{1}).
Then, the kernel requests the CPU to allocate new physical
EPC pages using the SGX instruction (i.e.,
\cc{EADD}~\cite{costan2016intel}) (\WC{2}).
This consequently makes CPU to create an EPCM entry corresponding to
\PAzero and \PAone, which contains \VAzero and \VAone, respectively~(\WC{3}).
At this point, in order to share and thus alias this EPC page later,
\sys additionally records the mappings for \VAzero and
\VAone~(\WC{4}).

When creating a new container, \sys aliases the previously allocated
EPC pages.
This entails to insert the recorded address mapping---i.e., (\VAzero,
\PAzero) and (\VAone, \PAone)---to the page table of the new
container~(\WC{5}).
Therefore, when the new container accesses the memory with \VAzero or
\VAone, the CPU would accordingly translate it to the aliased EPC page
located at \PAzero or \PAone using (i) the page table of the new
container and (ii) the corresponding EPCM entry.

From the performance perspective, sharing the enclave's memory significantly
reduces the latency since it avoids copying, verifying and encrypting the
initial memory contents for every new confidential container.
Furthermore, it reduces the memory footprint by enabling to
share the common data between two containers, avoiding to
duplicate the same data in each enclave every time.

\begin{figure}[t]
  \centering
    \includegraphics[width=\linewidth]{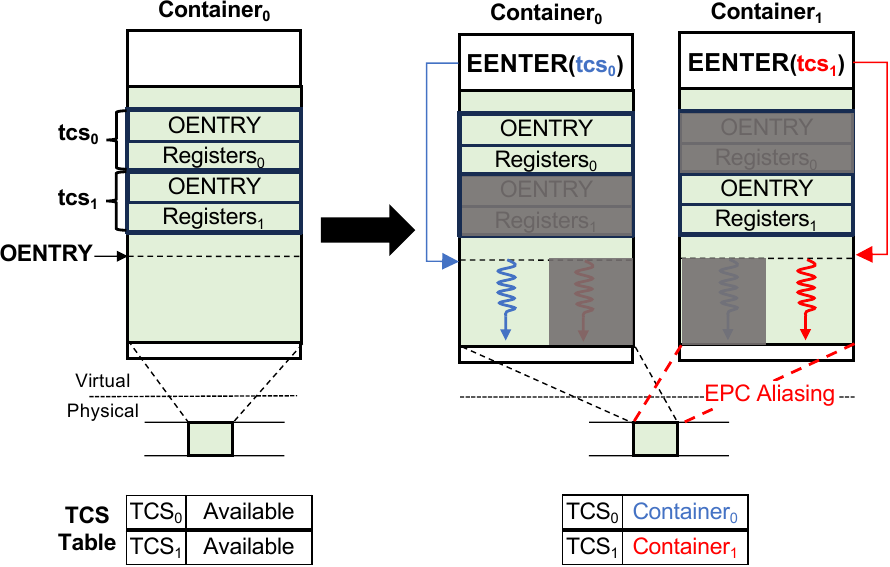}
    \caption{Constructing container-independent enclave threads}
  \label{fig:design-TCS-workflow-order-old}
\end{figure}

\subsection{Assigning Enclave Threads to Each Container}
\label{ss:design-thread}

Then, \sys provides the thread abstraction for the host
kernel as if each process is running its own thread in the
dedicated enclave.
To be specific, \sys assigns the threads in the same
enclave to each different process so that the host kernel
can run each enclave thread in different containerized
environments---e.g., providing different file systems using
namespace~\cite{namespace}.
One may wonder it is a vulnerable design as the host kernel
can illegally change the container environment for a given
enclave thread, but we can guarantee the security by
carefully designing a sanity check logic in the enclave,
which is explained in~\autoref{s:security}.
In the following, we elaborate the details how we achieve
the thread abstraction in the context of Intel SGX~\cite{gcore-cc}.

\PP{Technical Analysis: Multi-threads in SGX Enclave}
In order to support parallelism, SGX designs unique schemes
for secure multi-threading within an enclave, which we refer
to as enclave threads~\cite{costan2016intel}.
In particular, the execution contexts of enclave threads are
managed by Thread Control Structures (TCSs), which are also
stored in the EPC pages to protect against adversarial
system components.
More technically, each TCS manages the entry point and the
CPU register context per enclave thread.
First, the entry point ensures that an enclave thread always
starts or resumes (i.e., \cc{EENTER} or \cc{ERESUME}) at the
code address designated in the given TCS (i.e., \cc{OENTRY}
field).
The entry point often contains security checks and
enforcement to sanitize the inputs from non-enclave
context.
Second, the CPU register context ensures that a paused
enclave thread is always resumed as expected.
This is carried out by saving (and restoring) all CPU
register values to (and from) the TCS\footnote{More
  precisely, TCS stores the reference to State Save Area
  (SSA) in the EPC pages.}, which happens when the enclave
thread exits (and resumes).
It is worth noting that initial TCS pages are measured as
other EPC pages, thereby preventing untrusted system components
from breaking its initial integrity.

The technical catch here is that unlike the typical use-cases of
multi-threading in SGX, we find that an enclave thread does not need
to be bound with a specific process.
In fact, an enclave thread can be migrated from one process to
another.
Specifically, when switching from the non-enclave execution context to
the enclave execution context, any TCS page can be selected and such a
selection does not restrict which process performs the switch (i.e.,
which process performs \cc{EENTER} or \cc{ERESUME}).
\sys leverages this operational property to enable individual
execution context per container while sharing the EPC memory pages.

\PP{Individual Enclave Thread per Container}
In order to support an enclave thread per container while aliasing EPC
pages, \sys designs management schemes to map between a TCS page and a
container.
As such, while sharing EPC pages, \sys assigns a dedicated
TCS page per enclave thread, where each enclave thread is
associated with a different containerized process.

More technically, \autoref{fig:design-TCS-workflow-order-old} shows
how \sys supports an individual enclave thread per container.
In this figure, we assume that a single container (i.e., $\cc{Container}_0$)
and an enclave were initially created before, where no enclave code has
been executed yet.
Accordingly, \sys initializes a TCS table, which indicates that all
TCS pages are available.
Next, a new container is created (i.e., $\cc{Container}_1$), which
shares the EPC pages through \enclaveAliasing.
To execute enclave threads per container, \sys first picks an
available TCS page per container (e.g., $\cc{tcs}_{0}$ for
$\cc{Container}_0$ and $\cc{tcs}_{1}$ for $\cc{Container}_1$,
respectively).
Then each container starts the execution of the enclave thread by
entering the enclave (i.e., \cc{EENTER}) with the chosen TCS
page.
Completing the execution of the enclave thread, the TCS table is
accordingly updated to mark which TCS page is now returned back to be
available.

\subsection{Ensuring Isolation Guarantees in an Enclave}
\label{ss:design-isolation}

The final step toward sharing an enclave is to isolate the
enclave's memory between the processes using it (i.e.,
enclave thread of each process).
In case of SGX enclaves, \sys cannot rely on page table
isolation~\cite{kpti}, as untrusted host kernel has the full
control of the paging structures.
Instead, we implement a software based memory isolation in
the enclave as shown in the previous
works~\cite{ahmad2021chancel,shen2020occlum,gu2022lightenclave,marcela2019enclavedom}---i.e.,
intra-enclave isolation.
However, we want to note that it is merely due to the design
of Intel SGX~\cite{gcore-cc}, and confidential VMs can
employ the paging mechanism that is securely implemented by
the trusted guest kernel in the VM.
We further explain how we implemented the isolation in SGX
enclaves for each application in~\autoref{s:eval}.

%% file: impl.tex
\section{Implementation}
\label{s:impl}

\sys's implementation consists of
i)~\sys controller, which is a composition of a tailored
system software stack (implemented in host kernel) and
userspace layer, and
ii)~\sys runtime, which manages the operations in the
enclave.

As \sys controller, we first modified Linux SGX
driver~\cite{sgxdriver} and Linux
kernel~\cite{linuxkernelv62} for the EPC aliasing
(i.e.,~\autoref{ss:design-memory}).
Especially, we implemented new \cc{ioctl} syscalls for
remembering the virtual to physical address mappings and
populating the same mappings in the other virtual address
space.
Based on it, we implemented the userspace layer to invoke
the \cc{ioctl} for remembering the mappings (by the process
that created an enclave), and populating the mappings (by
the other process that wants to use the enclave).
For the enclave threads, we implemented Linux SGX driver to
bookkeep which process uses which TCS page
(i.e.,~\autoref{ss:design-thread}).
Then, we made the userspace layer to get the pointer for
corresponding TCS page through \cc{ioctl} and enter into the
enclave with it.
We implemented \sys controller with 490 LoC in Linux SGX
driver, 10 LoC in Linux kernel, and 250 LoC in Gramine
LibOS~\cite{gramine}.
It is worth noting that we do not protect \sys controller,
and its integrity does not affect the security of the
components in enclave (i.e., explained more
in~\autoref{s:security}).

For \sys runtime which runs in the enclave, we modified
Gramine LibOS~\cite{gramine} to support real world applications
without modification.
To be specific, we reused most of the components in Gramine
LibOS, but enabled it to identify the process (based on a
requested TCS page) such that the integrity of the files
(outside the enclave) can be checked per process.
On top of it, we ran actual applications (i.e.,
Node.js~\cite{nodejs} for confidential serverless computing, and
Redis~\cite{redis} for confidential database) while they
internally isolate the threads in the enclave.
We elaborate more on the intra-enclave isolation for each
application in~\autoref{s:eval}.
We implemented 1800 LoC of Gramine LibOS for \sys runtime,
170 LoC in Node.js for confidential serverless computing,
and 270 LoC in Redis for confidential database.

%% file: eval.tex
\section{Evaluations on Applications}
\label{s:eval}

This section evaluates the performance improvement of \sys
on confidential serverless computing, and confidential
database.
We first describe the evaluation setup
(\autoref{ss:eval-setup}), then introduce the evaluation
results of each application in the following
(\autoref{ss:eval-serverless} and
\autoref{ss:eval-database}).

\subsection{Evaluation Setup}
\label{ss:eval-setup}

We evaluated \sys on 64-core Intel Xeon Gold 6348 CPU
machine which supports SGX2 feature~\cite{xeon}.
Especially, we ran all the experiments in a QEMU virtual
machine~\cite{qemu} with 160GB memory and 48GB EPC size,
which runs a Linux kernel 6.2.0~\cite{linuxkernelv62}.

\PP{Models for Comparison}
In order to clearly demonstrate the performance improvement
of \sys, we compare three models (for each application) as
summarized below:
\begin{itemize}[leftmargin=*]
  \setlength{\itemsep}{0.1pt}
  \item \native represents the conventional application that
        does not employ confidential computing (i.e., using
        only the container).
        This model shows the maximum performance that \sys
        can achieve.

  \item \strawman represents the secured version of the
        application that employs state-of-the-art
        confidential container~\cite{arnautov2016scone,CNCFCOCO}---i.e., each
        container creates its own SGX enclave.
        Especially, we used Gramine LibOS~\cite{tsai2017graphene} as the SGX
        runtime, and ran the bare-metal application on it.

  \item \sys implements our design such that the same
        SGX enclave is shared across different containers.

\end{itemize}

\subsection{Confidential Serverless Computing}
\label{ss:eval-serverless}

For the confidential serverless computing, we used
OpenWhisk~\cite{openwhisk}, which is a real-world serverless
computing platform widely used for analyzing the
performance.
In particular, for every request, OpenWhisk creates a new
container which includes the Node.js~\cite{nodejs} runtime, and
runs the serverless function implemented using
JavaScript (i.e.,~\native).
Then, the function receives the user's data, runs the code
on it, and returns the result.
In this respect, we implemented \strawman model to create a
new container, and a new SGX enclave in it, which loads the
Node.js runtime~\cite{nodejs} (on the Gramine LibOS~\cite{tsai2017graphene}) to
securely run the functions.

\PP{Intra-Enclave Isolation for Confidential Serverless
  Computing}
In order to integrate \sys with OpenWhisk, we further
employed V8 Isolate~\cite{v8isolate} to isolate the
functions in the same enclave.
To be specific, for each request, \sys's OpenWhisk creates a
new container (as usual), aliases an enclave into the new
address space, and runs the function using a new enclave
thread that is sandboxed by V8 Isolate (i.e., \sys).
Thus, V8 Isolate ensures the threads cannot access each
other's memory, and we can guarantee the containers using
the same enclave are isolated (as long as the implementation
of V8 Isolate is trusted).

\begin{table}[t]
  \centering
  \captionof{table}{List of evaluated serverless
    functions.}
  \input{fig/eval/table-workload}
  \label{tab:eval-functions}
\end{table}

\PP{Benchmark Functions}
We performed the evaluation with 9 serverless functions
written in JavaScript as shown
in~\autoref{tab:eval-functions}.
We selected the functions from SeBS~\cite{marcin2021sebs} and Google
Sunspider~\cite{sunspider}, which cover a wide range of serverless
functions that potentially receive sensitive data as input.

\begin{figure}[t]
  \centering
  \includegraphics[width=\linewidth]{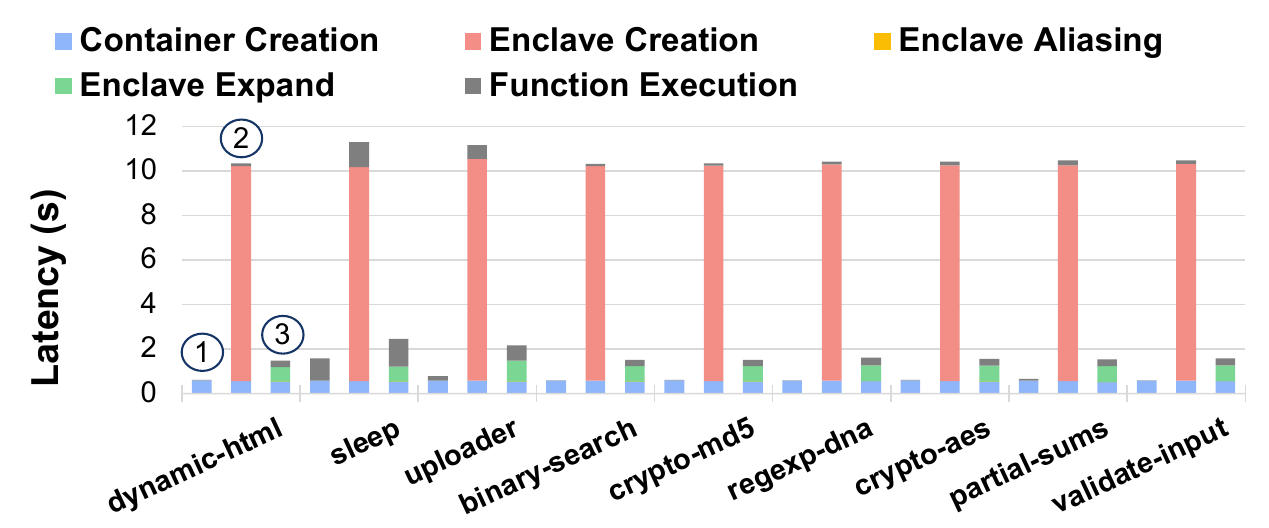}
  \caption{Breakdown of latency for running a function.
    \protect\WC{1}: \native,
    \protect\WC{2}: \strawman,
    \protect\WC{3}: \sys.
    "Enclave Creation" includes the time for initializing
    the enclave and loading software components in it.
    "Enclave Expand" denotes the time to allocate new EPC
    pages to accommodate new function instance. }
  \label{fig:eval-serverless-latency}
\end{figure}

\subsubsection{Performance Improvement of Startup Latency}
\label{sss:eval-serverless-latency}
\PP{Single Latency}
We measured the average latency for running each
serverless function as shown in~\autoref{fig:eval-serverless-latency}.
\sys showed 4.54-6.98\x latency speedup over the \strawman
model as it eliminates the overheads to create a new enclave
for each function instance (i.e., red portion in the bars).
On the other hand, the latency of \enclaveAliasing (which is
added instead of the enclave creation) was significantly
low, ranging from 2.32 to 3.01 milliseconds (i.e., less than
1\% of the entire latency).
\strawman model consumes almost 10 seconds to initialize an
enclave, and load Gramine LibOS and Node.js into it every
time (i.e., ``Enclave Creation'').
However, \sys does not impose such overheads, and only needs
to allocate new EPC pages for the new function instance,
which takes about 27.4-45.6\% of the entire latency.

\begin{figure}[t]
  \centering
  \includegraphics[width=\linewidth]{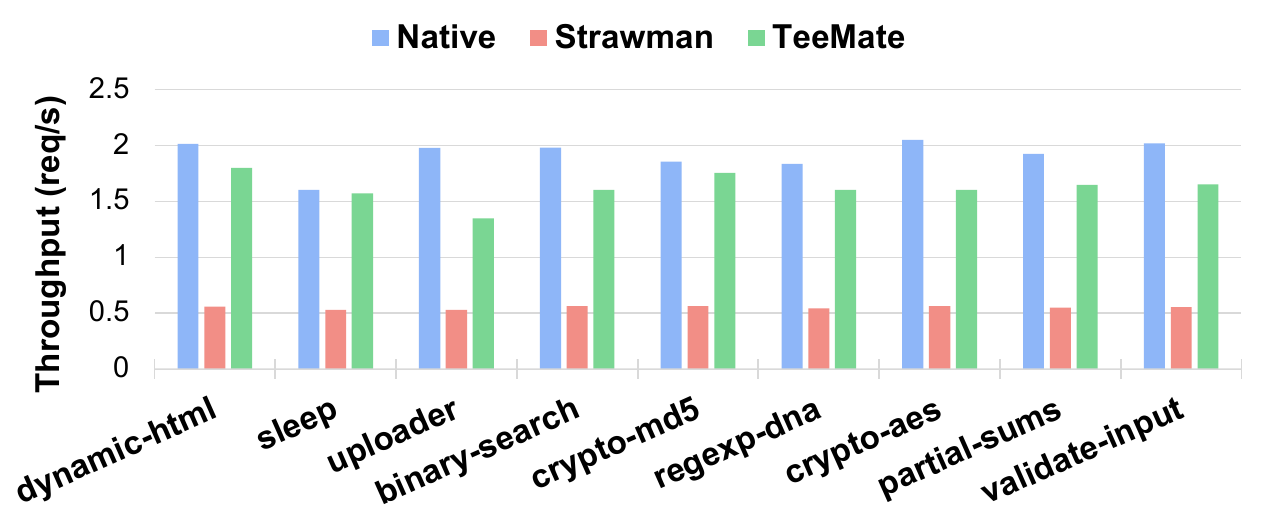}
  \caption{Throughput when 8 concurrent requests are handled.}
  \label{fig:eval-serverless-throughput-8}
\end{figure}
\begin{figure}[t]
  \centering
  \includegraphics[width=\linewidth]{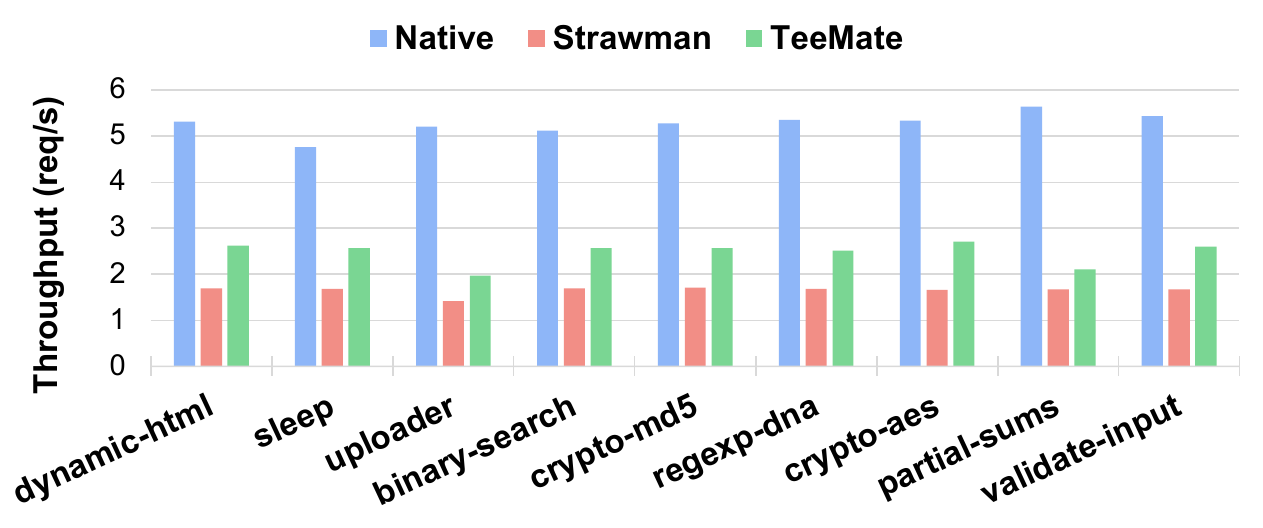}
  \caption{Throughput when 64 concurrent requests are handled.}
  \label{fig:eval-serverless-throughput-64}
\end{figure}
\PP{Throughput}
We evaluated the throughput improvement of \sys by invoking
a burst of requests simultaneously and observing the total
time to complete all the requests---i.e., throughput is
computed as the number of requests divided by the completion
time.
To this end, we invoked 8 and 64 requests respectively, for
each model with each function.
As a result, \sys exhibited 1.26-3.21\x higher throughput
than \strawman as illustrated
in~\autoref{fig:eval-serverless-throughput-8} (i.e., 8
requests invoked simultaneously),
and~\autoref{fig:eval-serverless-throughput-64} (i.e., 64
requests).
Even when compared to \native model, \sys showed comparable
performance by decreasing only 5.5-62\% of the native
throughput.

One thing to note is that the throughput gain of \sys
decreases as the number of requests increases from 8 to 64.
We suspect it is because the SGX driver internally reserves
a lock~\cite{sgxdriver} when allocating the EPC pages, and
the lock contention increases as the number of concurrent
requests increases.
In order to relieve this lock contention, \sys may employ
further optimizations such as batch processing the EPC
allocation requests~\cite{orenbach2017eleos}, or reusing the
allocated EPC pages.
In addition, we want to note that \sys can also create more
enclaves for the same function to maximize the throughput,
while we used only one enclave for this evaluation.

\begin{figure}[t]
  \centering
  \includegraphics[width=\linewidth]{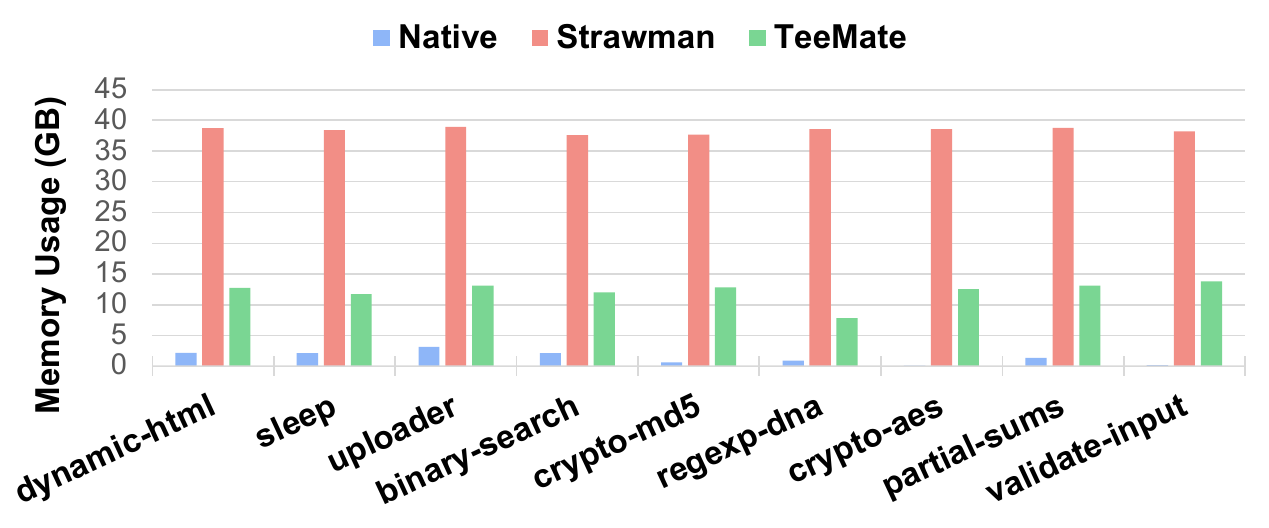}
  \caption{Peak EPC memory usage when 64 concurrent requests are handled.}
  \label{fig:eval-serverless-memory}
\end{figure}

\subsubsection{Performance Improvement of Memory Footprint}
\label{sss:eval-serverless-memory}
In order to evaluate the memory efficiency of \sys, we
compared the peak memory usage of \native, \strawman, and
\sys when a bunch of requests are received.
For each function, we invoked 64 concurrent requests
simultaneously and measured the peak memory usage.

As illustrated in~\autoref{fig:eval-serverless-memory}, \sys
showed 2.8-5\x lower memory usage compared to \strawman.
This is because \strawman model needs to load a new runtime
(i.e., Gramine library OS~\cite{tsai2017graphene} and
Node.js~\cite{nodejs}) within each enclave, while \sys needs
to create only a lightweight V8 Isolate~\cite{v8isolate} on
top of the shared runtime (in the shared enclave).

On the other hand, \sys uses more memory (i.e., 207MB) than \strawman
(i.e., 114MB) when handling only one request.
This is because \sys needs additional memory to accommodate the isolation
mechanisms (i.e., V8 Isolate).

\subsection{Confidential Database}
\label{ss:eval-database}

For the confidential database, we used Redis~\cite{redis}, which
is an in-memory database widely serviced by cloud
platforms~\cite{kim2019shieldstore}.
Specifically, Redis uses fork-based snapshot~\cite{redisFork} to
support data persistence.
To be specific, Redis process periodically forks a child
which performs the snapshot by writing the database into the
storage, while the parent continues to handle the requests
(i.e.,~\native).
Copy-on-write semantics~\cite{Copy-on-write} are well suited for this
mechanism as the full page copy is performed only when a
write request is received to the parent.

However, running Redis on state-of-the-art confidential
containers cannot employ the copy-on-write semantics as the
parent's enclave and child's enclave cannot share the
memory (i.e.,~\strawman).
To be specific, if we run the Redis on Gramine LibOS~\cite{tsai2017graphene}
as usual, fork from the parent creates a new child process
(outside the enclave) and creates a new enclave again, then
Gramine LibOS would copy all the memory contents from the
parent to the child enclave to preserve the semantics of
fork---i.e., child process inherits the same memory space as
the parent.
Thus, the entire pages are copied on fork without supporting
copy-on-write.

\PP{Intra-Enclave Isolation for Confidential Database}
In order to integrate \sys with Redis, we implemented
software address translation~\cite{orenbach2017eleos} in the Redis, which
isolates memory accesses by each process.
Especially, after a child process is forked, two threads
from the parent and the child still run using the same
addresses (in the same enclave with different page tables).
However, when a write request is received, we made Redis to
copy the target page and execute the request on the copied
page (i.e., copy-on-write) such that the other process
cannot see the updated contents (i.e.,~\sys).
This ensures the memory isolation as the two processes
access exactly same physical pages as long as their contents
are the same (which does not harm any security guarantee).

\begin{figure}[t]
  \centering
  \begin{subfigure}{0.47\columnwidth}
    \includegraphics[width=\linewidth]{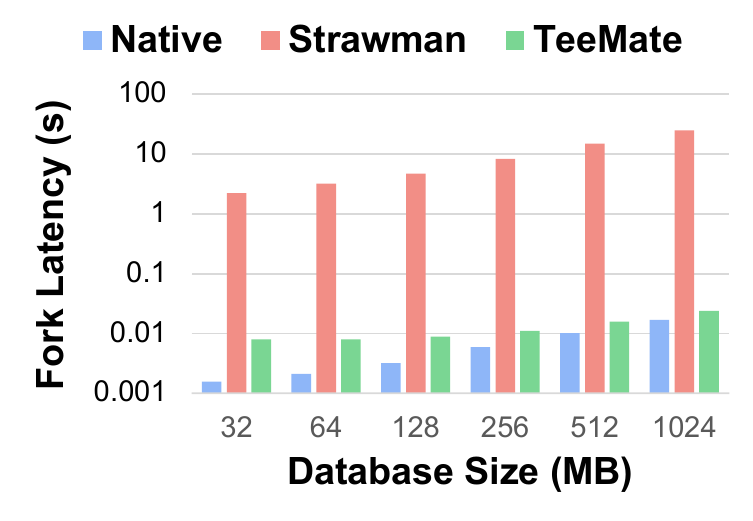}
    \caption{Latency for forking a child.}
    \label{fig:eval-db-latency}
  \end{subfigure}
  \begin{subfigure}{0.47\columnwidth}
    \includegraphics[width=\linewidth]{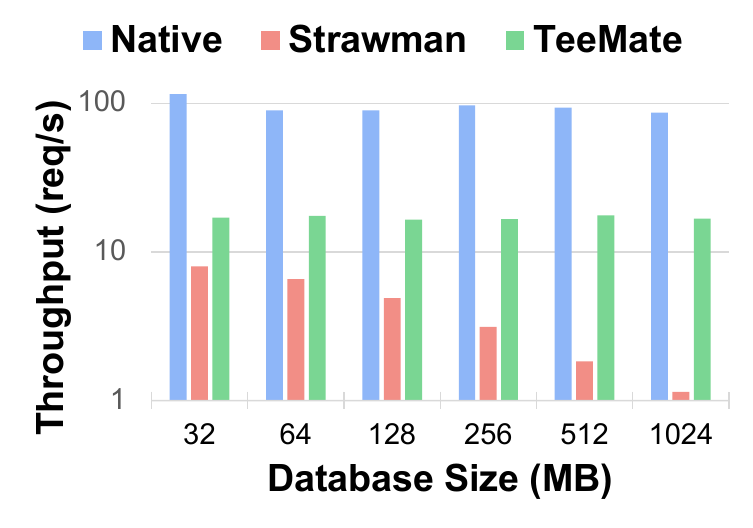}
    \caption{Throughput.}
    \label{fig:eval-db-throughput}
  \end{subfigure}
  \caption{Latency evaluation of confidential database.}
  \label{fig:eval-db-performance}
\end{figure}

\subsubsection{Performance Improvement of Fork Latency}
\label{sss:eval-database-latency}
Since the fork-based snapshot necessarily forks a child
process, we measured that latency of the Redis process with
different size of database.
As shown in \autoref{fig:eval-db-performance}-(a), \sys
showed 277.6-1046.6\x latency speedup compared to \strawman
model.
This is because \sys does not create a new enclave and copy
the parent enclave's memory to the child, but the two
processes can share the memory.
Fork latency in \sys consists of the latency to create a new
enclave thread and makes a child process to execute the
enclave thread, which is much faster than enclave creation
and copying the entire memory.
In addition, \sys takes the advantages of copy-on-write by
which it copies the page only when a write request is
received to the parent.

\PP{Throughput}
We also evaluated the throughput improvement of \sys by
invoking a burst of requests while the database performs
fork-based snapshot.
As a result, \sys exhibited 2.1-14.6\x higher throughput
than \strawman as illustrated
in~\autoref{fig:eval-db-throughput}.
Since Redis cannot handle any request during the fork system
call is handled, the throughput of \strawman was
significantly affected due to the longer fork latency.
On the other hand, \sys was able to achieve better
throughput thanks to the short fork latency and
copy-on-write operations.

\subsubsection{Performance Improvement of Memory Footprint}
\label{sss:eval-database-memory}
\begin{figure}[t]
  \centering
  \begin{subfigure}{0.47\columnwidth}
    \includegraphics[width=\linewidth]{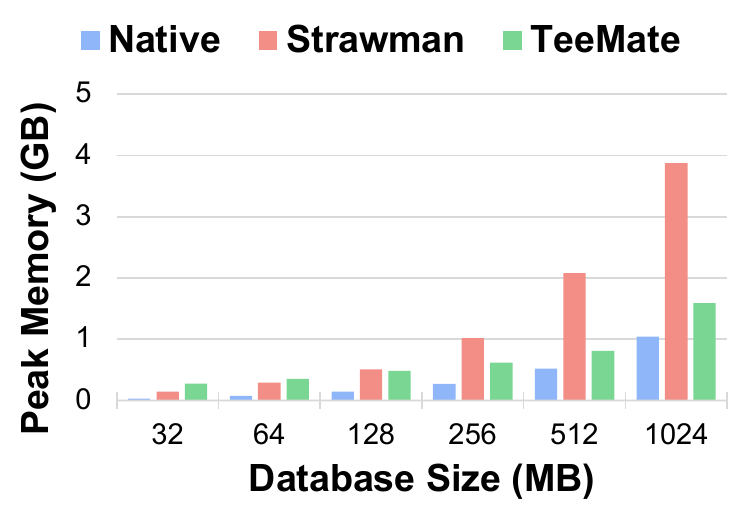}
    \caption{Peak memory usage}
    \label{fig:eval-db-memory}
  \end{subfigure}
  \begin{subfigure}{0.47\columnwidth}
    \includegraphics[width=\linewidth]{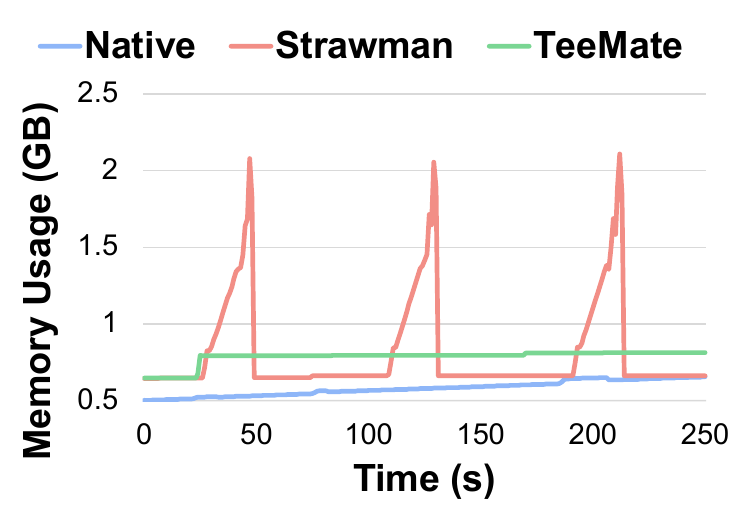}
    \caption{Overall memory usage}
    \label{fig:eval-db-macro}
  \end{subfigure}
  \caption{Memory usage of each model}
  \label{fig:eval-db}
\end{figure}
We evaluated the memory efficiency of \sys by measuring the
peak memory usage and overall memory usage of the Redis
while it runs as usual and also performing fork-based
snapshots.
As illustrated in \autoref{fig:eval-db}-(a), \sys showed
much lower peak memory usage compared to \strawman model as
the database size increases.
While \strawman model suffers from large memory footprint
due to the duplicated memory of the parent and child
enclaves, \sys avoids such issues thanks to the shared
enclave, showing 41\% lower memory usage at maximum.

Meanwhile, when the database size is small, \sys showed
slightly larger memory usage because it needs additional
memory for software address translation.
In addition, we observed the memory overhead incurred by the
heap allocator of Gramine LibOS when creating a new
thread~\cite{gramine}.

We also evaluated the overall memory usage of each model as
described in~\autoref{fig:eval-db}-(b).
For this evaluation, we set the database size to be 512MB
and generated random requests to the database engine.
Additionally, we set the database engine to create a
snapshot once every minute, which is the same as default
configuration of Redis~\cite{redisFork}.
As expected, \sys showed significantly lower memory usage
than \strawman model when performing fork-based snapshot
operations.
However, in normal situations when not performing a
snapshot, \sys showed slightly higher memory usage due to
the aforementioned reasons.

%% file: fig/eval/table-workload.tex
\resizebox{0.8\linewidth}{!}{
  \begin{tabular}{l|l}
    \cmidrule[1pt]{1-2}
    Name & Description \\
    \cmidrule[1pt]{1-2}
    dynamic-html & creating html using input \\
    sleep & sleep for 1 second \\
    uploader & upload local file to remote storage \\
    binary-search & binary tree search using input as a key \\
    crypto-aes & AES encryption/decryption \\
    crypto-md5 & MD5 hash computation \\
    partial-sums & input array summation \\
    regexp-dna & DNA sequence processing \\
    validate-input & input string processing \\
    \bottomrule
  \end{tabular}
}

%% file: secanalysis.tex
\section{Security Analysis}
\label{s:security}

In this section, we analyze the security guarantees of \sys.

\PP{Security Guarantees of Sharing an Enclave across
  Multiple Processes}
While it seems unsafe at first glance to share a single
enclave across multiple processes, it does not compromise
the security guarantees of confidential 
computing~\cite{Confidential-computing}.
The key question is that ``what if the compromised host
enters the enclave in the context of another process
(different from the one that initially created the
enclave)?''.
This is the same question as ``what if the compromised host
modifies all the states of a process (except the memory of
the enclave) to be the other one, and enters the enclave?''.
The second question is already the common threat model of
confidential computing, and widely discussed in the
academia~\cite{gcore-cc,sev,inteltdx,armcca}.
The answer is that ``the host cannot compromise the enclave
as all the security critical data and logic should be
located in the enclave''---i.e., the security guarantees of
the confidential computing (and those of \sys) still hold.

The key is that the code in an enclave should not believe
any information passed over from the outside enclave.
In other words, all the security critical data should be
managed in the enclave, which includes the identity of the
client who invoked the request, TLS encryption key, and the
hash of the files to be checked for integrity.
All the security operations such as decrypting the
ciphertext from users, or checking the integrity of opened
file should be performed in the enclave also, which is the
same as the conventional use-cases~\cite{tsai2017graphene}.
For example, if a compromised host runs an enclave thread in
a different containerized environment (that was not supposed
to be used), then, access to a security critical, but
different data should be detected in the enclave.

\noindent\textbf{Security Limitations of \sys.}
Compared to the previous approaches that protect the
workloads of each process using each dedicated
enclave~\cite{feng2021scalable,li2021confidential,gu2022lightenclave,shixuan2023reusableenclave}, 
\sys has weaker isolation guarantees as it employs intra-enclave isolation.
The intra-enclave isolation should be implemented on
software (e.g., V8 Isolate~\cite{v8isolate}) in case of Intel SGX,
while it can be implemented using the paging mechanism in
case of confidential VMs (e.g., AMD SEV~\cite{sev}, and Intel
TDX~\cite{inteltdx}).
However, we believe \sys still achieves the major goal of
confidential computing---i.e., removing cloud providers from
the trusted path.
Furthermore, we want to note that software fault isolation
(that \sys should use in case of Intel SGX) is already
widely used in various real world scenarios (e.g., WASM
sandbox in browser~\cite{wasm-sandbox}, kernel model
sandboxing~\cite{mckee2022preventing}).
While these may have vulnerability, \sys can benefit from
ongoing researches to improve the software fault isolation.

%% file: discuss.tex
\section{Discussion}
\label{s:discuss}

\subsection{Applicability to VM based Confidential Computing}
The core observation of \sys is that an enclave does not
have to be bound to a process as the enclave is just a set
of physical resources while the process is the abstraction
managed by host kernel.
This does not depend on the type of enclave, and it is also
applied to the VM based confidential computing (e.g., AMD
SEV~\cite{sev}, and Intel TDX~\cite{inteltdx}).
This is because the VM is also a set of physical resources
(i.e., virtual memory and virtual CPU).
While Linux kernel manages a single VM by a single
process~\cite{kvm}, the VM can be shared as long as the memory
and thread abstractions are preserved.

In this respect, we demonstrate that design primitives of
\sys can be applied to AMD SEV-SNP VMs.
Specifically, we check that multiple processes can share the
same enclave pages (i.e., secure pages in AMD terminology)
of a single SEV-SNP VM.
In the following, we provide a brief explanation of access
control and address translation mechanism of SEV-SNP, and
then describe how we leverage these mechanisms to adopt the
design of \sys in SEV-SNP.

\PP{Technical Analysis of SEV-SNP: Access Control and Address Validation Mechanism.}
ASID (Address Space Identifier) is a crucial element of access control
mechanism in AMD SEV-SNP~\cite{sevsnp}.
The ASID serves as a unique identifier for each VM, and it is used to select a
VM encryption key (VEK).
VEK is used for encrypting and decrypting enclave pages.
Thus, only the VM with the correct ASID can access the enclave pages encrypted
with the corresponding VEK. 
Importantly, the ASID can be modified at runtime by the hypervisor during a
VMEXIT event~\cite{mengyuan2021crossline}.

SEV-SNP employs a reverse map table to validate the address translation when a
VM accesses an enclave page.
When a VM allocates new enclave page, the reverse map table records the VM's
ASID and the guest physical address of the page.
When the VM later accesses the enclave page, the hardware checks whether the
VM's ASID and the guest physical address match the corresponding record in the
reverse map table.
This prevents malicious hypervisor from launching a page remapping attack,
similar to purpose of address translation mechanism in SGX (mentioned
in~\autoref{ss:design-memory}).

\PP{Enclave Page Aliasing in SEV-SNP}
Based on above analysis, we propose a method to alias enclave pages in
SEV-SNP by matching two key pieces of information: i)~ASID, and ii)~guest
physical address.
To share the same enclave pages (i.e., secure pages of the
same VM) between multiple processes, the hypervisor should
first record the ASID of the VM which allocates the pages.
Then, when another process attempts to access the page, the
hypervisor should assign the ASID of the VM that originally
allocated the page.
Furthermore, the hypervisor needs to assign the same nested
page table to ensure that the enclave page is accessed using
the same guest physical address.

We verify through experiments that our method actually works in SEV-SNP machine.
We modified Linux KVM module to change ASID and nested page table pointer (nCR3)
of VM during VMEXIT event to those of the VM that originally allocated the page.  
We will also investigate the applicability of \sys to other confidential VM
technologies such as Intel TDX~\cite{inteltdx}, and ARM CCA~\cite{armcca} in the
future work.

\subsection{Use-cases of \sys}
\sys enables multiple processes (that are managed by
untrusted host kernel) to efficiently share the data within
an enclave.
Thus, this feature can be widely applied to the confidential
computing use-cases that need frequent data communications,
while the resources are managed by untrusted host kernel
(e.g., micro-service architecture using Kubernetes~\cite{kubernetes},
or big data analysis using Spark~\cite{spark}).
However, it should be carefully applied as this approach
necessarily bloats the trusted computing base (TCB) by
implementing the control logic in the enclave.

%% file: relwk.tex
\section{Related work}
\label{s:relwk}
\PP{Confidential Container}
Confidential container is gaining popularity due to its ability to meet the
needs of efficient resource management by cloud providers and data protection by
cloud users using TEE (Trusted Execution Environment).
SCONE~\cite{arnautov2016scone} is one of the first confidential container
system, which integrates SGX enclave~\cite{gcore-cc} to Docker
container~\cite{docker}.
Additionally, the Cloud Native Computing Foundation's Confidential Container
project~\cite{CNCFCOCO} is actively conducting various research related to
confidential containers.
For example, the project implements both process-level containers using Intel
SGX~\cite{cocoenclavecc} and microVM-level
containers~\cite{cococonfidentialcontainer} using confidential virtual machines
like Intel TDX~\cite{inteltdx} and AMD SEV~\cite{sev}.
TZ-Container~\cite{zhichao2021tzcontainer} utilizes ARM TrustZone to create a secure execution environment for each container process.
Although many confidential container system have been proposed, they have not
challenged to the universal misconception that only one process can use a
specific enclave.

\PP{Confidential Serverless Computing}
Incorporating TEE with serverless computing is gaining more interests as it
provides strong security guarantees to protect sensitive data and code even in a
compromised environment.
Clemmys~\cite{trach2019clemmys} uses SGX enclave to block
platform provider from introspecting the memory, and devised
a cryptographic model to prevent maliciously modifying the
order of function chaining from the platform provider.
AccTEE~\cite{goltzsche2019acctee} and S-FaaS~\cite{fritz2019sfaas} introduced a
fair and trustworthy resource accounting for confidential serverless computing.
SEVeriFast~\cite{holmes2024severifast} implemented new bootstrap scheme in AMD SEV~\cite{sev} for low startup
latency in VM-based confidential serverless computing. 
Reusable enclave~\cite{shixuan2023reusableenclave} achieved low startup latency
in confidential serverless computing by secure enclave reset mechanism.
PIE~\cite{li2021confidential} extends the SGX design through
hardware modification to optimize the startup latency and
function chaining latency by memory sharing between enclaves.
Several works have also provided the ground for sharing
memory between the
enclaves~\cite{jason2022elasticlave,lee2022cerberus}, but
all of them need to modify the hardware, and are not able to
be applied to current platforms.
It is worth noting that the memory sharing approaches
mentioned above focus on sharing memory between different
enclaves, whereas \sys focuses on sharing a single enclave
across multiple containers.

\PP{Confidential Database}
Confidential database protects database engine on untrusted cloud so that the
confidentiality and integrity of data and queries are guaranteed. 
For instance, EnclaveDB~\cite{christian2018enclavedb} used SGX enclave to protect all database state
including the data and query from the cloud provider. 
Library OS for Intel SGX such as Graphene-SGX~\cite{tsai2017graphene} and
Haven~\cite{Baumann2014haven} are also used to run confidential database without
application modification.
OBLIVIATE~\cite{ahmad2018obliviate} proposed data oblivious filesystem to
prevent side-channel attacks against malicious cloud provider toward database
engine.

\PP{Intra-Enclave Isolation}
For the performance and security issues, several previous
works have suggested to split the enclave into multiple
isolated regions.
Chancel~\cite{ahmad2021chancel} protects the client's data
in application, which handles each client's request using
dedicated thread.
Chancel uses a compiler-based per-thread isolation inside
enclave.
Occlum~\cite{shen2020occlum} enables multi-process
applications to run efficiently inside enclave by software
fault isolated processes, leveraging Intel MPX and code
instrumentation.
These isolation approaches can also be used for the
isolation mechanism of \sys, but we want to note that V8
Isolate is more appropriate for isolating high-level
programming languages that are widely used in serverless
computing (e.g., JavaScript, and Python).
LightEnclave~\cite{gu2022lightenclave} and
EnclaveDom~\cite{marcela2019enclavedom} leveraged Intel MPK
for fine-grained isolation inside enclave.
However, it needs hardware modification under the threat
model of confidential computing as Intel MPK basically
depends on host kernel.

%% file: conclusion.tex
\section{Conclusion}
\label{s:conclusion}
This paper proposes \sys, which introduces a new approach to
utilize the enclaves in the host's perspective.
Especially, we found there is a universal misconception that
an enclave must be dedicated to a single process that
created it, and we break this assumption by sharing an
enclave across multiple processes.
To this end, we design the primitives to preserve the memory
and thread abstraction for a single SGX enclave to be shared
across multiple processes.
Based on it, we implemented confidential serverless
computing and confidential database, and demonstrated that
\sys shows significant latency speedup and memory usage
reduction compared to the same applications using
state-of-the-art confidential containers.